\documentclass[aps,letterpaper,amssymb,twocolumn,pra]{revtex4}
\usepackage{amssymb}
\usepackage[usenames,dvips]{color}
\usepackage{graphicx}
\usepackage{amsmath}
\usepackage{times}
\usepackage{subfigure}
\begin{document}
\title{ Symmetry-Breaking and Symmetry-Restoring Dynamics of a Mixture of 
Bose-Einstein Condensates in a Double Well}
\author{Indubala I Satija and Philip Naudus (George Mason University, Fairfax, VA), Radha Balakrishnan ( Institute of Mathematical Sciences,
Chennai, India),
Jeffrey Heward ,Mark Edwards( Department of Physics, Georgia Southern University,
Statesboro, GA ) and
and Charles W Clark
(National Institute of Standards and Technology, 
Gaithersburg, Maryland )}
\begin{abstract}
{We study the coherent nonlinear tunneling dynamics of a binary mixture 
of Bose-Einstein condensates in a double-well potential. We demonstrate 
the existence of a new type of mode associated with the "swapping" of the two species in the two wells of the potential. In contrast
to the symmetry breaking macroscopic quantum self-trapping (MQST) solutions, the swapping modes correspond
to the tunneling dynamics that preserves the symmetry of the double well potential.
As a consequence of two distinct types of 
broken symmetry MQST phases where the two species localize in the different potential welils or coexist in the same well,
the corresponding symmetry restoring swapping modes result in dynamics where the
the two species either avoid or chase each other.
In view of the possibility to control the interaction between the species,
the binary mixture offers a very robust system to observe these novel 
effects as well as the phenomena of Josephson oscillations and pi-modes}.
\end{abstract}

%\pacs{03.65.Vf, 05.45.Mt, 05.45.Ac}
\maketitle

\section{Introduction}
\label{intro}

Ultracold laboratories have had great success in creating Bose-Einstein 
condensates (BECs)~\cite{ande} in a variety of atomic gases such as Rubidium 
(Rb), Lithium (Li), Sodium (Na) and  Ytterbium (Yb). These quantum fluids 
exist in various isotopic forms as well as in different hyperfine states.  
The rapid pace of development in this field has led to condensates which 
are robust and relatively easy to manipulate experimentally.  In particular, 
the tunability of inter-species and intra-species interactions~\cite{thal} 
via magnetic and optical Feshbach resonances makes the BEC mixture a very 
attractive candidate for exploring new phenomena involving quantum coherence 
and nonlinearity in a multicomponent system.  

The subject of this paper is to investigate the tunneling dynamics of a 
binary mixture of BECs in a double well potential. A single species of 
BEC in a double well is called a bosonic Josephson junction (BJJ), since 
it is a bosonic analog of the well known superconducting Josephson junction.
In addition to Josephson oscillations (JO), the BJJ exhibits various novel 
phenomena such as $\pi$-modes and macroscopic quantum self-trapping (MQST), 
as predicted theoretically~\cite{smer,milb}. In the JO and the $\pi$-modes,
the condensate oscillates symmetrically  about the two wells of the potential.
In contrast to this, the MQST dynamics represents a broken symmetry phase as the tunneling solutions 
exhibit population imbalance between the two wells of the potential. These various features have been 
observed experimentally~\cite{exptbjj}.  Our motivation is to explore whether 
new phenomena arise when there are two interacting condensates trapped in a 
symmetric double well.

Although our formulation and results are valid for a variety of BEC mixtures, 
our main focus here is the Rb family of two isotopes, namely the mixture of 
$^{87}$Rb and $^{85}$Rb, motivated by the experimental setup at JILA\cite{thesis}.
The scattering length of $^{87}$Rb is known to be $100$ 
atomic units while the interspecies scattering length is $213$ atomic units.
In experiments, the scattering length of $^{85}$Rb can be tuned using the 
Feshbach resonance method~\cite{feshbach_reference}.

The ability to tune the scattering length of one of the species makes 
this mixture of isotopes an ideal candidate for studying the coupled BJJ 
system.  First, it opens up the possibility of exploring the parameter space 
where the Rb 85--85 scattering length is equal to the Rb 87--87 scattering 
length.  As will be discussed below, this symmetric parameter regime 
simplifies the theoretical analysis of the system and also captures most of 
the new phenomena that underlie the dynamics of the binary mixture.  
Furthermore, the tunability of the $^{85}$Rb scattering length can be 
exploited to study a unique possibility where one of the species has a 
negative scattering length, a case which strongly favors the $\pi$-mode 
oscillations that have not been observed so far.

In our exploration of nonlinear tunneling dynamics of coupled BJJ systems, 
the MQST states are found to be of two types.  In the broken--symmetry MQST 
state, the two components may localize in different wells resulting in a 
phase separation or they may localize in the same well and hence coexist.
By varying the parameters such as initial conditions,
the phase--separated broken--symmetry MQST 
states can be transformed to a symmetry--restoring phase where the species
continually ``avoid" each other by swapping places between the two wells.
In other words, if the dynamics is initiated with both species in the same potential well,
the sustained tunneling oscillations are seen where the two species swap places between the well one and the well two.
From the coexisting MQST phase, one can achieve symmetry restoring swapping dynamics by
initiating the dynamics with two species in the separate wells.
In this case, the emergence of the swapping modes can be 
interpreted as a phase where the two species ``chase" each other.  

The paper is organized as follows. In section \ref{two_mode}, we discuss 
the model and use the two--mode approximation to the Gross--Pitaevskii (GP) 
equation to map it to a system of two coupled pendulums with momentum--%
dependent lengths and coupling.  Section \ref{fixed_pts} discusses the 
stationary solutions and their stability.  These results enable us to look 
for various qualitatively different effects without actually solving the GP 
equations.  Section \ref{dynamics} describes the numerical solutions of the 
GP equations as various parameters of the system are tuned. Although we have
explored the multi-dimensional
parameter space, the
novelties attributed to the binary mixture in a double well trap are presented in a restricted parameter space where
the scattering lengths of the two species are equal. Additionally, in our numerical results described here,
we fix the ratio of Rb 87--87 interaction to Rb 85-87 interaction to be $2.13$.
This restricted parameter space is accessible in the JILA setup and provides a simple means to describe various
highlights of the mixture dynamics. Section \ref{experiment} provides additional details of the JILA setup relevant
for our investigation.
  A summary is 
given in section \ref{summary}.

\section {Two-mode GP Equation for the Binary Mixture}
\label{two_mode}

In the semiclassical regime where the fluctuations around the mean values 
are small, the two-component BEC is described by the following coupled
GP equations for the two condensate wave functions $\Phi_l(x,t)$, with 
$l= a,b$ representing the two species in the mixture. 

\begin{eqnarray*}
i \hbar  \dot{\Phi}_a &=& (-\frac{\hbar^2}{2m_a} \nabla^2+V_a)
\Phi_a + (g_a |\Phi_a|^2 +g_{ab}|\Phi_b|^2) \Phi_a\\
i \hbar \dot{\Phi}_b &=& (-\frac{\hbar^2}{2m_b} \nabla^2+V_b)
\Phi_b + (g_b |\Phi_b|^2 +g_{ab}|\Phi_a|^2) \Phi_b.
\label{gpe}
\end{eqnarray*}

Here, $m_l$, $V_l$ and $g_l=4\pi \hbar^2 a_l/m_{l}$, denote respectively,
the mass, the trapping potential and the  intra-atomic
 interaction of each species, with
 $a_l$ as the corresponding scattering length.
$g_{ab} = 2\pi \hbar^2 ( 1/m_a+1/m_b) a _{ab} $ is the inter-species
interaction,  where $a_{ab}$  is the corresponding scattering length.
For the JILA experiment, in view of the tight confinement
of the condensate transverse to the trap, it is sufficient to consider
the corresponding one-dimensional GPE equations.

The condensate wave functions satisfy the normalization conditions,
\begin{equation}
\int d^3 r | \Phi_l |^2 = N_l
\label{norm}
\end{equation}
The total number of atoms in the mixture is $N = N_{a} + N_{b}$.
In the weakly linked limit, the dynamical oscillations  of the
 two-component BEC can be described by
two wave functions representing the  condensate in each trap labeled by
$k = 1, 2$,  with the  spatial and the temporal
contribution factored as follows:

\begin{eqnarray}
\left( \begin{array}{cc} \Phi_a \\ \Phi_b \end{array} \right)=
\left( \begin{array}{cc} \chi^a_1(x) \psi^a_1(t)\\ \chi^b_1(x) \psi^b_1(t) \end{array}\right)
+\left( \begin{array}{cc} \chi^a_2(x) \psi^a_2(t)\\ \chi^b_2(x) \psi^b_2(t)
\end{array}\right)
\label{wf}
\end{eqnarray}

The localized spatial modes $\chi^{(l)}_{k}(x)$  are computed as sums and differences
of the symmetric and antisymmetric solutions of the time--independent,
coupled GP equations.

To derive the equations of motion in the two--mode approximation, we
introduce $z_{l}(t)$, the population imbalance, and $\phi_{l}(t)$, the relative phase
of of species $l$ between 
the left and right sides of the double well potential,
\begin{equation}
z_l\left(t\right)=(|\psi^l_1|^2-|\psi^l_2|^2)/N,
\label{zl}
\end{equation}

\begin{equation}
\phi_l\left(t\right)=(\theta^l_1-\theta^l_2).
\label{phil}
\end{equation}

where $\psi^{(l)}_{k}(t)=|\psi^{(l)}_{k}(t)|\exp{i\theta^{(l)}_{k}}$
are the time--dependent coefficients in the two--mode equations
.Substituting equations (\ref{wf}) into the coupled 
GP equations and integrating over the spatial degrees of freedom, we obtain 
the following four coupled, nonlinear, ordinary differential equations which
we refer to as the ``two--mode'' model.
\begin{eqnarray}
\dot{Z}_a&=&-\bar{K}_a \sqrt{1-Z_a^2}\,\, \sin\phi_a\\
\dot{Z}_b&=&-\bar{K}_b \sqrt{1-Z_b^2} \,\,\sin\phi_b\\
\dot{\phi}_a&=&\bar{\Lambda}_a f_a Z_a+\Lambda_{ab} f_b Z_b 
+ \bar{K}_a \frac{Z_a}{\sqrt{1-Z_a^2}} \,\,\cos \phi_a\\
\dot{\phi}_b&=&\bar{\Lambda}_b f_b Z_b+\Lambda_{ab} f_a Z_a 
+ \bar{K}_b \frac{Z_b}{\sqrt{1-Z_b^2}} \,\,\cos \phi_b.
\label{tuneqn}
\end{eqnarray}
where $Z_l=z_l/f_l$.
In the above equations, $f_l=N_l/N$ denotes the fraction of atoms of 
species $l$, while $ \bar{K}_l$ and $\bar{\Lambda}_{l}$ are given by
\begin{eqnarray*}
\bar{K}_{a} &=& 
K_{a}-2f_a C_{a}\sqrt{1-Z_{a}^{2}}\cos\phi_{a} +
f_b D_{ab} \sqrt{1-Z_{b}^{2}}\cos\phi_{b}\\
\bar{K}_{b} &=& 
K_{b}-2 f_b C_{b} \sqrt{1-Z_{b}^{2}}\cos\phi_{b} +
f_a D_{ba} \sqrt{1-Z_{a}^{2}}\cos\phi_{a}\\
\bar{\Lambda}_a &=& 
\Lambda_a+C_a\\
\bar{\Lambda}_b &=& 
\Lambda_b+C_b
\end{eqnarray*}
In the above, the space and time-independent parameters $K_{l}$, $\Lambda_{l}$, $\Lambda_{ab}$, 
$C_{l}$ and $D_{ab}$ can be expressed in terms of various microscopic parameters that appear in GP equation and
the localized modes,
$\chi^{l}_{k}(x)$ and their overlap (integrated over spatial degrees of freedom).
The explicit expressions for these 
parameters are given in the Appendix.
 
The parameters $K_{l}$ describe the tunneling amplitude while $\Lambda_{l}$
is related to the corresponding scattering length of the species. The 
parameters $C_{l}$ and $D_{ab}$ have their origin in the overlaps between 
the spatial modes $\chi_{1}$ and $\chi_{2}$, and are expected to be small in the weak tunneling 
limit.  These overlaps modify the bare parameters denoted by the interaction 
$\Lambda_{l}$ and the tunneling $K_{l}$.  Consequently, we have a variable 
tunneling model, since the tunneling parameters $\bar{K}_{l}$ depend 
explicitly on the dynamical variables $Z_{l}$ and $\phi_{l}$.

In  our analysis, we will mostly restrict ourselves to the case where the two 
species are equally populated, namely $f_a=f_b=1/2$.  
In this case, the 
above system of equations (\ref{tuneqn}) can be
viewed as the Hamilton equations
in terms of the canonical 
variables $Z_{l}$ (momenta) and $\phi_{l}$
(co-ordinates), with the Hamiltonian given by the following form:
\begin{equation}
H=\frac{1}{2}[\bar{\Lambda}_a Z_a^2+\bar{\Lambda}_b Z_b^2
+2\Lambda_{ab} Z_a Z_b]-\sum_{l=a,b} \bar{K}_l
\sqrt{1-Z_l^2}\,\,\cos \phi_l .
\end{equation} 

For the case where the overlap between the spatial modes $\chi_1$ and $\chi_2$ can be neglected,
and the effective tunneling $\bar{K}_{l}$ can be replaced by its bare value $K_{l}$
the above system can be viewed as a coupled pair of non-rigid pendulums,
with momentum--dependent lengths. The coupling between the pendulums 
is also momentum dependent.

We parenthetically remark that  this system can also be mapped to a pair
of classical spins with Cartesian components
\begin{eqnarray*}
S^l_x&=&\sqrt{1-Z_l^2}\,\, \cos\, \phi_{l}\\
S^l_y&=&\sqrt{1-Z_l^2}\,\, \sin\,\theta_{l}\\
S^l_z&=&Z_l,
\end{eqnarray*}
so that $(S^l)^2=1$. Thus the spin vector locates a point on the unit
sphere given by polar angles $\theta_{l},\phi_{l}$,  with
$Z_l=\cos\theta_{l}$.  The corresponding spin Hamiltonian, written in terms of bare variables,
can be shown to be
\begin{eqnarray*}
H=\sum_{l=a,b} [\frac{1}{2}(\Lambda_l+ C_l) (S^l_z)^2+ C_l (S^l_x)^2-
K_l S^l_x] \\
+\Lambda_{ab} (S^a_z S^b_z)-D_{ab} (S^a_x S^b_x).
\end{eqnarray*}

The spin mapping provides an alternative means to visualize the effective
interaction between the two species during the tunneling.
If we ignore the spatial overlap integrals between the localized modes in two wells, ( $C_l=0$, $D_{ab}=0$ ),
the binary mixture of condensates in two-mode approximation, maps to two Ising-like spins in a transverse
magnetic field. The full two-mode variable tunneling  feature induces
XY-like spin interaction.

In this paper, we find it convenient to exploit mapping
to the coupled pendulums, for exploring tunneling dynamics in the coupled
BJJ. Although we have explored the full two-mode variable tunneling model, we will only discuss the constant
tunneling case ( $\bar{K}_{l}$ replaced by $K_{l}$ and $\bar{\Lambda}_l$ replaced by $\Lambda_l$.),
as the overlap integrals are small and the various novel effects of the mixture described here are found to
be robust and unaffected by the variable tunneling parameters.

\section{ Stationary Solutions: Fixed Points }
\label{fixed_pts}

The solutions of the coupled system are characterized by the interactions 
$\Lambda_l$, the ratio of the tunneling amplitude for the two species, 
$K_{a}/K_{b}$ which we denote by $R$ as well as the initial phase difference 
$\phi_l(t=0)$ and the initial population imbalance $Z_l(t=0)$.  In the multi--
dimensional parameter space the equilibrium or fixed--point solutions, in 
which the right--hand-sides of Eqs.\ (\ref{ftun})-(\ref{tun}) are zero, 
provide an effective tool to classify different categories of behavior of 
the system.

In general, these fixed--point equations are transcendental and have to be 
solved numerically. 
However, in the symmetric case where $\Lambda_a=\Lambda_b=\Lambda$,
, $K_a=K_b=K$, the fixed point equations can be tackled analytically.
Further, as can be seen from Eqs.\ (\ref{ftun})--(\ref{tun}), the parameter 
$K$ can be eliminated in this case by rescaling $t$ ($t \rightarrow Kt$) 
and redefining  $\Lambda_x$ as $\Lambda_x \rightarrow \Lambda_x/K$.  Our detailed 
analysis shows that this special case captures many relevant phenomena 
characterizing the binary mixture in a double well. In this case, the fixed 
points belong to two broad categories as stated below, resulting in two types 
of small amplitude oscillations about these two fixed points.  It is important 
to note that this type of MQST phase does not exist in a BJJ with a single species.

\ \\
{\bf(I) Zero-mode Fixed Points} ($\phi^*_a=\phi^*_b=0)$\\
\ \\
(1)  $Z^{\ast}_a = Z^*_b = 0$
\ \\
(2)  $Z^*_a = -Z^*_b = \pm \frac{\sqrt{ (\Lambda_{ab}-\Lambda)^2-4K^2)}}
{|(\Lambda_{ab}-\Lambda)|}$
\bigskip\par
{\bf(II) $\pi$-mode Fixed Points } ( $\phi^*_a=\phi^*_b=\pi$ )\\
\ \\
(1) $Z^*_a = Z^*_b = 0$
\ \\
(2) $Z^*_a = Z^*_b = \pm \frac{\sqrt{ (\Lambda_{ab}+\Lambda)^2-4K^2)}}
{|(\Lambda_{ab}+\Lambda)|}$
\\
It should be noted that the mixed-mode Fixed Points, ($\phi^*_a=0$ and $\phi^*_b=\pi)$
, ($Z^*_a = Z^*_b = 0$)\\
are unstable for the restricted parameter regime we 
are considering here and hence will not be discussed.

The small oscillations about the fixed point ($Z^{\ast}=0,\phi^{\ast}=0$) result in Zero-mode while
small oscillations about ($Z^{\ast}=0,\phi^{\ast}=\pi$)  lead to $\pi$-mode. The oscillation frequencies are in the next subsection.
 
The non--trivial fixed points ($Z^{\ast}\ne 0$ ) result 
in solutions with population imbalance and lead to tunneling dynamics with macroscopic quantum self-trapping or the MQST. 
In view of the $a-b$ symmetry, we have two  sets of stationary solutions: $Z_x^{\ast}$ and $-Z_x^{\ast}$), ($x=a,b$) 
This suggests the possibility of modes 
where each species oscillates about the binary fixed points, going back and forth between the two wells.
Unlike MQST, these modes will preserve the symmetry of the double well. However,
in contrast to Zero-modes, these modes are non-linear and give rise to "swapping phase" that will be discussed later.

The emergence of fixed points with 
opposite signs for the two species, ($Z_a^{\ast} = -Z_b^{\ast}$) in the Zero mode phase suggests that MQST in 
Zero--mode is accompanied by phase separation of the two species. In contrast, 
in the $\pi$-mode MQST phase , the two species could coexist in the same potential well as ($Z_a^{\ast} = Z_b^{\ast}$). 
Therefore, the fixed point equations suggest
that $\pi$-modes mimic attractive interaction between the two species.

The onset from 
oscillatory to MQST phase corresponds to the values of the parameters where 
the non--trivial fixed points move from the complex to the real plane.
Alternatively, the condition for the broken symmetry phase can be obtained 
by linear stability analysis of the fixed--point equations. This is discussed 
in the next sub-section.

In the asymmetric case when $\Lambda_a \ne \Lambda_b$
the fixed points are obtained by solving the 
coupled transcendental equations:
\begin{eqnarray*}
(-1)^p \frac{K_a Z^{*}_a}{\sqrt{1-(Z^*_a)^2}}+
\frac{1}{2}(\Lambda Z^{*}_a + \Lambda_{ab} Z^{*}_b) & = &0\\
(-1)^p \frac{ K_b Z^{*}_b}{\sqrt(1-(Z^{*}_b)^2)}+
\frac{1}{2}(\Lambda Z^{*}_b + \Lambda_{ab} Z^{*}_a )& = &0,
\end{eqnarray*}
where $p=0(1)$ for $\phi^*_a=0(\pi)$ and $\phi^*_b=0(\pi)$..
\bigskip\par\noindent
Analogous to the symmetric case, both the Zero and the $\pi$-mode solutions including those corresponding to MQST can be found
numerically. As expected, for the MQST fixed points $Z^*_a \ne -Z^*_b$ in the Zero-mode and $Z^*_a \ne Z^*_b$ in the $\pi$-mode
and we do not have the permutation symmetry or the
$a-b$ symmetry.
However, unlike the symmetric case, $K_l$s  do not scale time $t$ 
and the parameters and hence the ratio $R=\frac{K_a}{K_b}$ 
emerges as a new parameter.

\subsection{ Normal Modes: Linear Stability Analysis of Fixed Points}
\ \\
Frequencies of small amplitude oscillations about 
about ($Z^*=0, \phi^*=0$) and ($Z^*=0, \phi^* = \pi$)
respectively referred to as the Zero-mode 
or the $\pi$-modes are given by
\begin{eqnarray*}
\omega^2=\frac{1}{2}
(K_a\Lambda^*_a+
 K_b\Lambda^*_b)+\frac{1}{2}
 \pm\sqrt{(K_a\Lambda^*_a-K_b\Lambda^*_b)^2+4K_a K_b\Lambda^2_{ab}} ]
\end{eqnarray*}
where
\begin{eqnarray*}
\Lambda^*_a &= &(-1)^p K_a+f_a\Lambda_a\\
\Lambda^*_b &= &(-1)^p K_b+f_b\Lambda_b,
\end{eqnarray*}

where $p=0$ for the Zero-mode , and $p=1$ for the $\pi$-mode.
In the symmetric case, with $\Lambda_a=\Lambda_b$ and 
$f_a=f_b$, the normal mode frequencies $\omega_0$ and $\omega_{\pi}$ simplify to,
\begin{eqnarray*}
\omega^2_0& =& K^2+K(\Lambda \pm \Lambda_{ab})/2\\ 
\omega^2_{\pi}& =& K^2-K(\Lambda \pm \Lambda_{ab})/2 
\end{eqnarray*}

The condition for the instability of the fixed point is determined when 
one of the normal mode frequencies become complex. This gives rise to new fixed points where $Z_x^* \ne 0$ resulting in
MQST phase where there is a population imbalance between the two wells of the double well potential for each species.
The condition for the existence of MQST is given by,
\begin{eqnarray}
f_a f_b \Lambda^2_{ab} &  \ge & \Lambda^*_a \Lambda^*_b
\end{eqnarray}

For the parameter
values where both the Zero and the $\pi$-modes coexist, $\pi$-mode frequencies
are smaller than the
Zero-mode frequencies.

\begin{figure}[htbp]
\includegraphics[width =1.0\linewidth,height=1.0\linewidth]{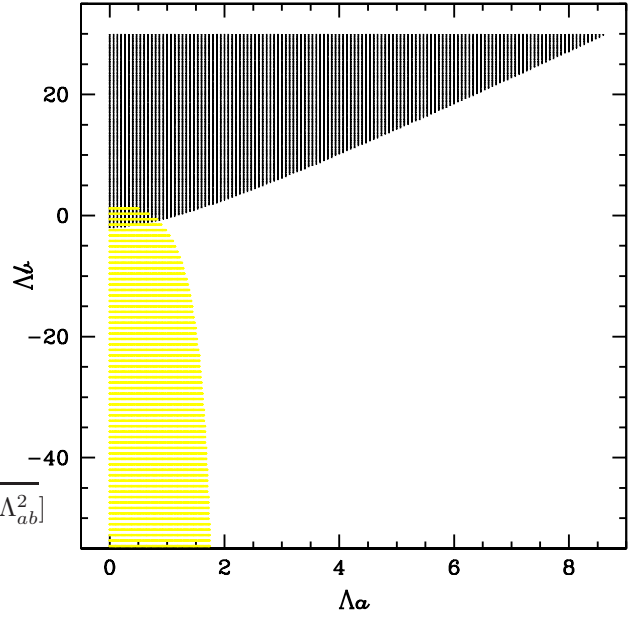}
% om.eps
\leavevmode \caption{(color online) 
The upper(black) and the lower(yellow) shaded regime 
 corresponds to the parameter values for the existence
of stable Zero-mode and $\pi$-mode with $R=1$. }
\label{fig1}
\end{figure}

Figure ~1 shows the values of $\Lambda_a, \Lambda_b$ where the tunneling 
is governed by the  Zero-mode and the
 $\pi$-mode.  For $\Lambda_b > 0$, the regime where the $\pi$-modes exist is is small but finite.
However, by tuning $\Lambda_b$ to negative values, the 
$\pi$-modes that have not been seen in earlier studies,
can be observed.
Variation with the parameter $R$, the tunneling ratio for the two species,
leads to similar results,
with the parameter space for the existence of $\pi$ mode increasing slightly with $R$.
The unshaded regime corresponds to
MQST phase.

\section{ Tunneling Dynamics with $\Lambda_a=\Lambda_b$ }
\label{dynamics}

We now describe numerical solution of the tunneling equations.
For small population imbalance, we confirm the dynamics predicted by the fixed points as discussed above.
However, numerical solutions also illustrate nonlinear modes, not described by the fixed point analysis.
The fact that new features continue to exist in the nonlinear regime, assures their robustness.

In our numerics, we set $\Lambda_{ab} = 2.13 \Lambda$ and study
the dynamics for different values of $\Lambda$. 
These conditions can be achieved by
first tuning the $g_b$ via a Feshbach resonance so that 
$\Lambda_a=\Lambda_b$.
The variation of $\Lambda$
corresponds to varying the number of atoms in the double well trap.
As already mentioned, $K$ can be eliminated by using  
$t \rightarrow Kt$ and 
$\Lambda \rightarrow \Lambda/K$.
The dynamics is governed by $\Lambda$ and the initial 
conditions: $Z_a(0)$, $Z_b(0)$, $\phi_a(0)$ and $\phi_b(0)$.

As we discuss below, tunneling solutions belong to three broad categories:\\
\bigskip\par
(I) " Zero-phase Mode " , characterized by $<\phi_l>=0$\\
\bigskip\par
(II) "$\pi$-phase Mode " characterized by $<\phi_l>= \pi$\\
\bigskip\par
(III) " Running-Phase Mode " characterized by $<\phi_l>$ proportional to $t$\\
\bigskip\par 

In the single species case, $<\phi_l>=0$ also corresponds to $<Z_l>=0$.
However, as we discuss below, in a binary mixture, we can have 
$<\phi_l>=0$ but $<Z_l> \ne 0$.
This gives rise to a broken symmetry MQST phase  in Zero-modes as well.

\subsection{ Zero-Modes }
\ \\

For $\Lambda < \Lambda^{0}_{c}\approx 1.77$, and $\phi_{a}(0)=\phi_{b}(0)=0$, 
and $|Z_l(0)| << 1$, both species execute small amplitude oscillations 
(like oscillations of a non-rigid pendulum) with,
 $<Z_l(t)>=0$ and 
$<\phi_l(t)>=0$ as shown in Fig. ~2. Such modes exhibit quasiperiodic dynamics characterized 
by superposition of sinusoidal modes with two competing frequencies.
As $Z_l(0)$ increases, we see large amplitude non-sinusoidal oscillations.
Therefore, in spite of the repulsive interaction between the two condensates, the two species execute a coherent 
oscillatory dynamics as expected from the zero-mode fixed point analysis described earlier.

\begin{figure}[htbp]
\includegraphics[width =1.0\linewidth,height=1.2\linewidth]{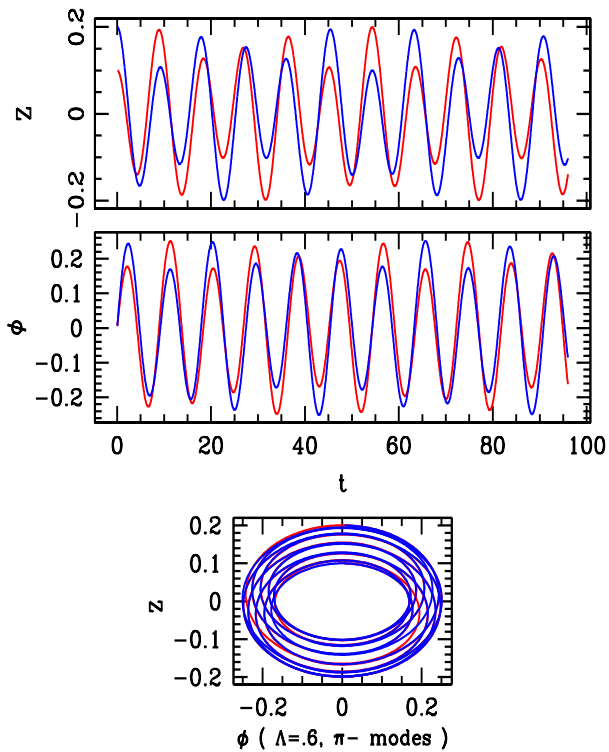}
%JOph0.eps
\leavevmode \caption{(color online) 
Time series for $Z_l$(top), $\phi_l$ (middle) and phase portrait
for $\Lambda=0.6$ with initial conditions shown in the figure. The red and blue corresponds to the $a$ and $b$
species. }
\label{fig2}
\end{figure}

\subsection { $\pi$-modes }

If the dynamics is initiated with $\phi_a(t=0)=\phi_b(t=0)=\pi$, both species 
oscillate in $\pi$-mode provided $\Lambda < \Lambda^{0}_{c}\approx 0.67$, 
and initial population imbalance is small ( $|Z_l(0)| << 1$) .  Analogous 
to the Zero-mode, the dynamics in the $\pi$ mode is in general quasiperiodic.
As seen in the figure, the motion is in phase with the slow mode and out of phase with the other.
Comparison with the Zero and the $\pi$-mode oscillations show that  species move more sluggishly in $\pi$-mode 
compared to the Zero-mode as the Zero-mode frequencies are larger than those of the $\pi$-mode.

\begin{figure}[htbp]
\includegraphics[width =1.0\linewidth,height=1.2\linewidth]{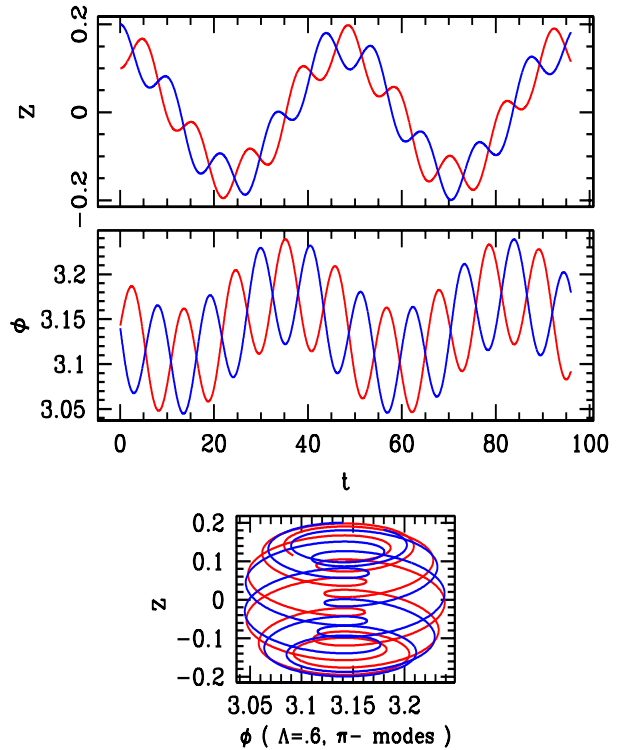}
%JOphpi.eps
\leavevmode 
\caption{(color online) Same parameters as Fig. ~2, the only exception 
being that $\phi_{l}(t=0)=\pi$ here.}
\label{fig3}
\end{figure}

\subsection{ Symmetry Breaking and Phase Separation: MQST in Zero-Mode }

\begin{figure}[htbp]
\includegraphics[width =1.0\linewidth,height=.7\linewidth]
{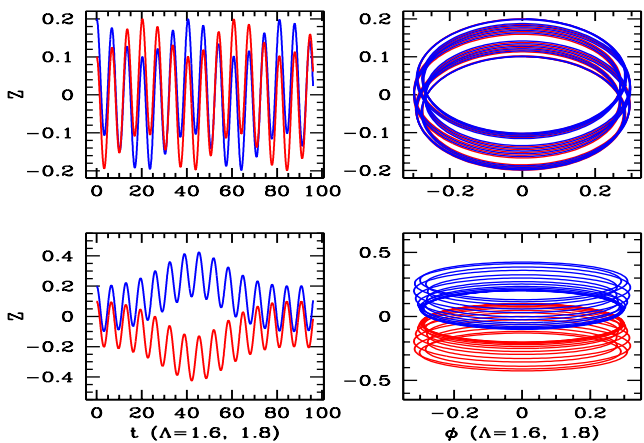}
%phase.separation.eps}
\leavevmode \caption{(color online) Transition from Josephson 
oscillations (top with $\Lambda=1.6$ ) to MQST with phase separation (bottom, with $\Lambda=1.8$ ),
obtained by varying $\Lambda$. The left and right plots 
 show the time series and phase portraits respectively.}
\label{fig4}
\end{figure}

Beyond a critical value of $\Lambda$, the system enters the symmetry breaking 
MQST phase, as predicted by the fixed point analysis earlier.
One of the novel aspects of the binary mixture is the existence 
of Zero-mode MQST accompanied by phase separation of the two species. 
Even with the initial conditions corresponding to both species abundance 
in the same well, the two components localize in the two different wells. 
In this case, transition to MQST is accompanied by phase separation:
although the two species overlap for some time, the  $<Z_a(t)>$ and the 
$<Z_b(t)>$ have opposite signs.

\subsection{ Symmetry Restoring and Phase Separation: Swapping-Mode }

As $\Lambda$ increases further,the system exhibits "swapping-modes" where 
the two species swap places between the two wells but remain phase separated as shown in Fig. ~4.
As seen in the figure (at $t=0$), the dynamics is initiated with positive population imbalance of both species.
However, the resulting dynamics corresponds to back and forth motion where the two species swap places between the
two wells. In contrast to MQST, the swapping dynamics restores the symmetry of the tunneling solution in the
double well. However, the two species remain mostly phase separated, {\it avoiding each other by swapping}.

In other 
words, the swapping phase is characterized by
 $<Z_a(t)>=<Z_b(t)=0$, but $<Z_a(t) Z_b(t) < 0>$ . That is, 
at a given instant of time, the two species are more likely to be found 
in the separate wells. 
Thus in the swapping mode, the two species
oscillate back and forth between the two wells and still manage to avoid each other.
The swapping is found to occur in the nonlinear Zero-mode as well as in the 
running mode . Furthermore, a transition from MQST to swapping 
phase can be achieved either by varying $\Lambda$ ( Fig. ~5) or by varying the initial 
conditions ( Fig. ~6).

\begin{figure}[htbp]
\includegraphics[width =1.0\linewidth,height=.7\linewidth]{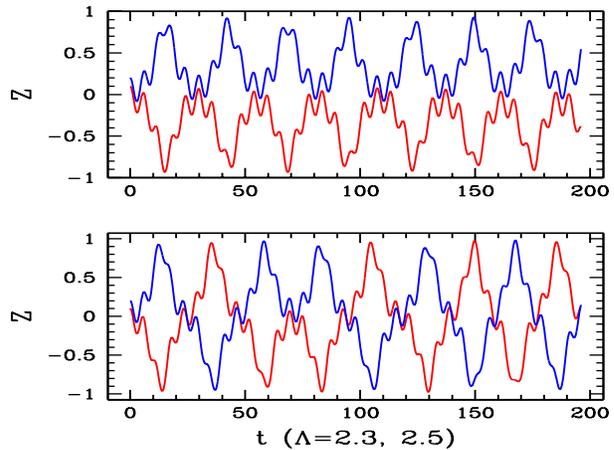}
%sw.eps
\leavevmode 
\caption{(color online) Symmetry restoring transition by varying 
$\Lambda$ where the upper panel with $\Lambda=2.3$ shows MQST phase with phase separation while the lower panel
with $\Lambda=2.5$ shows phase separation
due to swapping mode.}
\label{fig5}
\end{figure}

\begin{figure}[htbp]
\includegraphics[width =1.0\linewidth,height=.7\linewidth]
{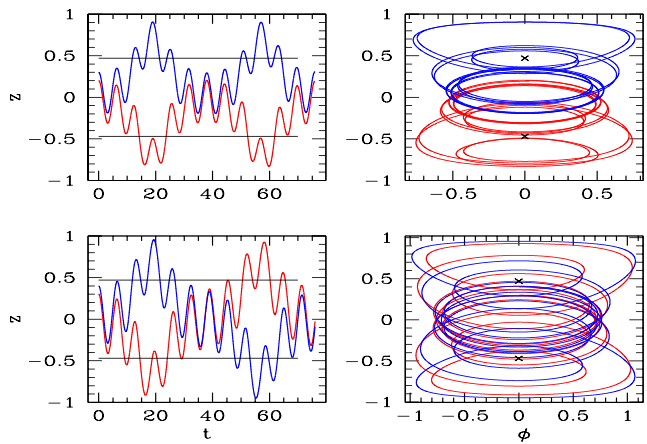}
%{swapinit.eps}
\leavevmode \caption{(color online) Symmetry restoring transition obtained by 
changing initial conditions ($Z_l(t=0)$) slightly for fixed $\Lambda=2$. Figure shows the time series as well as
the phase portraits where the straight line and crosses show the corresponding fixed points. }
\label{fig6}\end{figure}

\subsection{ Symmetry Breaking in $\pi$-Modes: Coexistence Phase }
\ \\

\begin{figure}[htbp]
\includegraphics[width =1.0\linewidth,height=.7\linewidth]{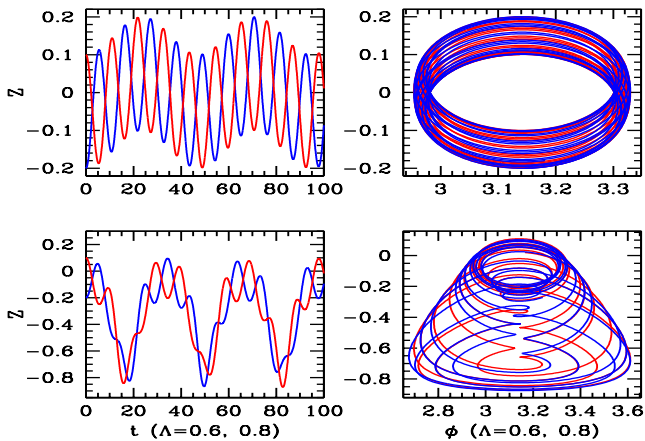}
%mqstpi.eps
\leavevmode 
\caption{(color online) Transition to MQST in $\pi$ modes. $\Lambda=0.6$ 
(top) and $\Lambda= 0.8$ (bottom) describe, respectively, the small 
amplitude $\pi$-mode oscillations and 
MQST in $\pi$-mode. }
\label{fig7}
\end{figure}

For $\Lambda < \Lambda^{\pi}_c \approx 0.67 $, and $\phi_a(t=0)=\phi_b(t=0)=\pi$, 
and $|Z_l(t=0)| << 1$,both species execute small amplitude oscillations 
with $<Z_l(t)>=0>$ and $<\phi_l(t)>=\pi$, as shown in Fig. ~7
Such modes are characterized by 
superposition of sinusoidal modes with two competing frequencies and the 
resulting dynamics is in general quasiperiodic.  As expected from the fixed 
point analysis, the two species with both inter and 
intra-species repulsive interaction can self-trap in the same well. That 
is , we have MQST where the species coexist in the same potential well, inspite of repulsive 
interaction among them.

\subsection{ Swapping in $\pi$-modes}
\ \\

\begin{figure}[htbp]
\includegraphics[width =1.1\linewidth,height=.8\linewidth]
{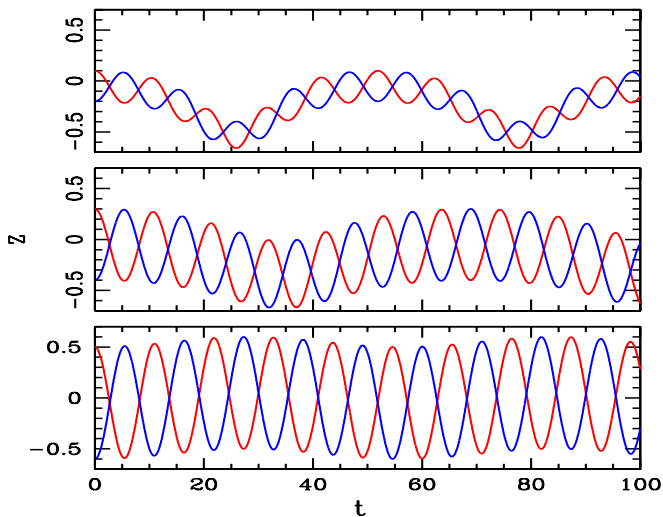}
%{pixt.eps}
\leavevmode 
\caption{(color online) Transition from broken symmetry (MQST in $\pi$ modes) 
to symmetric configurations, obtained by changing the initial population 
imbalance.  The three plots correspond to three different initial conditions.}
\label{fig8}
\end{figure}

As illustrated in figure \ref{fig8}, within the $\pi$-mode phase,
if the dynamics of the two species 
is initiated in separate wells, that is, $Z_a(t=0)$ and $Z_b(t=0)$ have 
opposite signs, the MQST phase can be destroyed when the initial 
population imbalance increases beyond a critical value.  The tunneling 
solutions become symmetric as MQST is replaced by swapping modes. In this 
case the swapping can be viewed as the two species "chasing" each other.

It should be noted that the swapping dynamics in the Zero and the $\pi$-modes is very
similar. However, swapping in the Zero-mode corresponds to two species avoiding each other while
swapping in the $\pi$-mode corresponds to one component chasing the other. This is because, in the Zero-mode,
species prefer residing in the separate wells while in the $\pi$-mode, they like to stay in the same well.
This
unique type of coherence between the two different species is one of the most fascinating
aspect of the binary mixture dynamics in double well potential.

\section{Experimental Realization}
\label{experiment}

The effects described in this paper should be realizable for condensate
mixtures that already exist in the laboratory.  One example in particular
is a mixture of $^{85}$Rb and $^{87}$Rb atoms that has been created in
several recent experiments at JILA~\cite{feshbach_reference, papp_exp}.
This system is relevant to the analysis in this paper because the scattering
length, $a_{85-85}$, that characterizes the interaction between $^{85}$Rb
atoms is tunable by an external magnetic field via a feshbach resonance
centered at approximately 155 Gauss~\cite{burke_and_bohn}.  Additionally,
the interspecies scattering length, $a_{85-87}$, is also tunable with two
feshbach resonances (for a $|2,-2\rangle_{85}$/$|1,-1\rangle_{87}$ collision)
located at approximately $B = 267$ Gauss and $B = 356$ Gauss.

In the most recent experiment~\cite{papp_exp}, a $^{85}$Rb/$^{87}$Rb BEC 
mixture was produced by trapping a thermal--gas sample of the mixture and 
performing evaporative cooling on the $^{87}$Rb which sympathetically cools 
the $^{85}$Rb.  The cold gas mixture is then transferred to an optical trap
that provides tight confinement transverse to the trapping beam and loose 
confinement along the beam.  If an additional pair of beams were applied
along this direction as was done in the Albiez experiment~\cite{exptbjj},
it would create a setup to which the analysis in this paper would apply.

\section{Summary}
\label{summary}

Existence of a variety of BEC species  
with tunable inter and 
intra--species scattering lengths makes BEC mixtures one of the most
attractive candidates for exploring novel phenomena involving quantum 
coherence and nonlinearity.  Our analysis, based on the two-mode GP 
equation for the two interacting species of BEC in a double well trap
unveils  a variety of phenomena describing broken symmetry as well as 
subsequent restoration of symmetry, as we change the parameters or the 
initial conditions.  Such coherence is found to exist over a broad range 
of parameters, establishing the robustness of the effects.

To make direct comparison with  experiments, we need to solve the coupled 
GP equations to obtain various parameters of the effective coupled 
pendulum system in terms of the microscopic parameters of the system and 
work in this direction is in progress.  Furthermore, by quantizing the 
Hamiltonian (coupled pendulum  or the spin Hamiltonian), we hope to study 
quantum dynamics of number fluctuations that may code the emergence of 
new quantum phases in the system.

\newpage

\appendix*
\section{Two--mode equation parameters}
\label{appendix}

With $\bar{g}_x=g_x N /\hbar$ ( $x=a,b, ab$ ), the various coupling constants in the
coupled equations can be shown to be given by,

\begin{eqnarray*}
\gamma_{a(b)}^{\pm}& =& \bar{g}_{a(b)} \int[ (\chi^{a(b)}_{\pm})^4] dr\\
\bar{\gamma}_{a(b)} &= & \bar{g}_{a(b)} \int[ (\chi^{a(b)}_+)^2 (\chi^{a(b)}_-)^2] dr\\
\Delta \gamma_{a(b)} &=& \gamma_{a(b)}^{-}-\gamma_{a(b)}^{+}\\
\Delta \gamma_{ab}& =& \bar{g}_{ab} \int[ (\chi^a_-\chi^b_-)^2-(\chi^a_+\chi^b_+)^2]dr\\
\Delta \bar{\gamma}_{ab}& =& \bar{g}_{ab} \int[ (\chi^a_-\chi^b_+)^2-(\chi^a_+\chi^b_-)^2]dr\\
\Lambda_a(b) &=& \bar{g}_{a(b)} \int[ 2(\chi^{a(b)}_+ \chi^{a(b)}_-)^2-1/4((\chi^{a(b)}_-)^2 -(\chi^{a(b)}_+)^2)^2)] dr\\
\Lambda_{ab} &=&2 \bar{g}_{ab} \int (\chi^a_+ \chi^a_- \chi^b_+ \chi^b_-)dr\\
\bar{K}_a& = &[\Delta E-f_a \Delta \gamma_a -f_b D_{ab}]/\hbar\\
\bar{K}_b& = &[\Delta E-f_b \Delta \gamma_b -f_a D_{ab}]/\hbar\\
C_a & = &  (\gamma^a_+ + \gamma^a_- -2\bar{\gamma_a})/2 \hbar\\
C_b & = &  (\gamma^b_+ + \gamma^b_- -2\bar{\gamma_b})/2 \hbar\\
D_{ab} &=& (\Delta \gamma_{ab}- \Delta \bar{\gamma}_{ab} )/2 \hbar
\end{eqnarray*}

For each species, the localized spatial modes $\chi^{(l)}_{\pm}(x)$ 
are obtained by adding and
subtracting the symmetric and the antisymmetric solutions of the time--independent coupled GPE equations.
The $\Delta E$ is the difference in the chemical potential between the (symmetric) ground and the
(anti-symmetric) first excited state of the coupled time-independent GPE equations.

\end{document}